# A novel fault localization with data refinement for hydroelectric units

Jialong Huang[1], Junlin Song[2], Penglong Lian[3], Mengjie Gan[4], Zhiheng Su[5], Benhao Wang[6],
Wenji Zhu[7], Xiaomin Pu[8], Jianxiao Zou[9], Shicai Fan[*]

*Abstract*—Due to the scarcity of fault samples and the complexity of non-linear and non-smooth characteristics data in hydroelectric units, most of the traditional hydroelectric unit fault localization methods are difficult to carry out accurate localization. To address these problems, a sparse autoencoder (SAE)-generative adversarial network (GAN)-wavelet noise reduction (WNR)-manifold-boosted deep learning (SG-WMBDL) based fault localization method for hydroelectric units is proposed. To overcome the data scarcity, a SAE is embedded into the GAN to generate more high-quality samples in the data generation module. Considering the signals involving non-linear and non-smooth characteristics, the improved WNR which combining both soft and hard thresholding and local linear embedding (LLE) are utilized to the data preprocessing module in order to reduce the noise and effectively capture the local features. In addition, to seek higher performance, the novel Adaptive Boost (AdaBoost) combined with multi deep learning is proposed to achieve accurate fault localization. The experimental results show that the SG-WMBDL can locate faults for hydroelectric units under a small number of fault samples with non-linear and non-smooth characteristics on higher precision and accuracy compared to other frontier methods, which verifies the effectiveness and practicality of the proposed method.

## I. INTRODUCTION

Hydroelectric unit is the key part of hydroelectric power generation, once the hydroelectric unit fails, it may lead to the paralysis of the whole power generating machine, and even irreversible disaster, so efficient and accurate methods of fault localization are particularly essential. Fault localization for hydroelectric unit plays an important role in the operation and maintenance which can reduce or eliminate the accident. It is of primary importance to provide the powerful methodology for the fault localization of hydroelectric unit not only of systems but also of data available [1].

There are two main difficulties in locating faults in hydroelectric units: Firstly, the fault data of hydroelectric units are usually scarce. Due to the low frequency of faults in hydroelectric units and the high cost of fault data collection and labeling, the number of available fault samples is limited, which is not enough to support the training and testing of effective fault localization models. To solve the problem of scarce fault data in fault diagnosis, few-shot learning (FSL) has been introduced to this area in recent years. The core issue of FSL is the unreliable empirical risk minimizer that makes FSL hard to learn [2]. Commonly used data augmentation models for small samples are GAN [3]-[4], meta-learning [5]-[6], and a combination of both [7].

Secondly, the fault data of hydroelectric units show nonlinearity and complexity. Characterization information of hydroelectric units is very important in the fault localization process. Currently, traditional methods for non-linear and non-smooth signal characteristics include those based on time and frequency domains [8]-[10], those based on the combination of frequency domain analysis and entropy theory [11]-[12]. However, since non-linear and non-smooth signals may contain complex dynamics, such traditional methods may involve complex calculations, which may affect their real-time performance. As deep learning has been maturing in recent years, there has been a surge in the number of applications of this method for fault localization such as the combination of deep learning algorithms and frequency domain analysis [13]-[14]. While deep learning captures the inherent characteristics of nonlinear and non-stationary signals better than traditional methods, it still requires a large amount of data to build accurate fault localization models. In addition, there are also applications in this field through ensemble learning [15]-[16], which can reduce the risk of overfitting a single model on training data by combining multiple models. However, due to the simplicity of the models and algorithms, the stability and accuracy depend on the model being integrated.

Combining the respective advantages mentioned above, in this paper, SG-WMBDL is proposed for fault localization of hydroelectric units. The SG-WMBDL includes three modules: data generation module and data preprocessing module and data localization module. Fault data generation module based on SAE and GAN model is proposed to connect the encoding part of SAE to the input of the generator and the decoding part of SAE to the output of the generator as an embedded mechanism embedded in an improved GAN network, which can effectively alleviate the problem of scarcity of fault data. Fault data preprocessing module based on improved wavelet noise reduction combining soft and hard thresholding and LLE method of manifold learning is effective noise reduction and dimensionality reduction of data. Fault localization module based on AdaBoost algorithm using dynamically adjusted weights based on category distribution combined with CNN, FCN deep learning can localize hydroelectric units effectively and perform further positive than pervious fault localization method. The main contributions and innovations of this paper are:

(1) This paper constructs a novel data generation module based on SAE and GAN that can effectively address the

Jialong Huang[1], Mengjie Gan[4], Benhao Wang[6], Jianxiao Zou[9] and Shicai Fan[*] are with Shenzhen Institute of Advanced Studies, University of Electronic Science and Technology of China, Shenzhen, Guangdong, China 518000.

Penglong Lian[3], Zhiheng Su[5] and Shicai Fan[*] are with School of Automation Engineering, University of Electronic Science and Technology of China, Chengdu, Sichuan, China 611731.

*Corresponding author (Email: shicaifan@uestc.edu.cn).

problem of scarcity of fault data.
(2) This paper proposes an improved wavelet noise reduction combining soft and hard thresholding and LLE method that can reduce the noise and effectively capture the local features of non-linear and non-smooth signals.
(3) This paper proposes the novel AdaBoost combined with deep learning that can achieve accurate fault localization.

The remainder of this paper is structured as follows. Section II introduces the basics of SAE, GAN, WNR, LLE, AdaBoost respectively. Section III introduces the overall architecture of the proposed SG-WMBDL method. The experimental data set and experimental results are given in Section IV. Finally, we conclude the conclusion in Section V.

## II. PRELIMINARIES

### A. Sparse Autoencoder

Sparse autoencoder (SAE) is an unsupervised feature learning neural network with three layers. The input layer that represents inputs, the hidden layer that represents learned features, and the output layer with the same dimension of the input layer that represents reconstruction[17]. The reconstruction error is defined as:

$$Loss = \frac{1}{2} \| \mathbf{Y} - \mathbf{X} \|^2 \quad (1)$$

where $\mathbf{X}$ is the input, $\mathbf{Y}$ is the prediction of SAE and $\|.\|$ represents the 2-norm of the vector.

SAE is based on the ordinary autoencoder added the sparsity penalty term, so that the neural network in the hidden layer of neurons in the case of more neurons can still be able to extract the features and structure of the sample. The sparsity penalty term is defined as $P_s$, the formula is as follow:

$$P_s = \alpha \sum_{c=1}^{e} \left( \rho \log \frac{\rho}{\hat{\rho}_c} + (1-\rho) \log \frac{1-\rho}{1-\hat{\rho}_c} \right) \quad (2)$$

where $\alpha$ is the sparsity penalty term parameter, $\rho$ is called the sparsity parameter which is an artificially given small value. $\hat{\rho}_c$ is the average activation value of the hidden unit c, which is defined as:

$$\hat{\rho}_c = \frac{1}{M} \sum_{d=1}^{M} h_c^d \quad (3)$$

where $M$ is the number of datasets, $h_c^d$ is the representation vector.

### B. Generative Adversarial Networks

Inspired by the two-player zero-sum game, generative adversarial networks (GAN) are composed of a generator and a discriminator, both trained with the adversarial learning mechanism. The aim of GAN is to estimate the potential distribution of existing data and generate new data samples from the same distribution[18]. GAN's training objectives are as follows:

$$\min_G \max_D L(D,G) = E_{x\sim P_d} \log[D(x)] + E_{x\sim P_n} \log[1-D(G(x))] \quad (4)$$

where $P_d$ is the real sample, and $D(x)$ is the probability of the real sample; $P_n$ refers to the generation of a copy of random noise $n$, and $D(G(x))$ is the probability that this generated sample is the real sample and the formula maximizes the discriminator D and minimizes the generator G.

### C. Wavelet Noise Reduction

The hydroelectric unit signal is decomposed into multiple wavelet coefficients using wavelet transform. The basis of the Fourier transform is swapped and the infinitely long trigonometric basis is replaced with a finite-length wavelet basis that decays, with the following formula:

$$WT(a,\tau) = \frac{1}{\sqrt{\beta}} \int_{-\infty}^{\infty} f(t) * \psi\left(\frac{t-\tau}{\beta}\right) dt \quad (5)$$

where $W$ denotes the wavelet transform of the signal $T_1$, the scale $\beta$ controls the expansion and contraction of the wavelet function, and the translation $\tau$ controls the translation of the wavelet function.

The soft thresholding method is to make the absolute value of the wavelet coefficients 0 when it is less than the given threshold, and to make them all minus the threshold when it is greater than the threshold, which is given in the following equation:

$$w_\lambda = \begin{cases} [\text{sgn}(w)](|w|-\lambda), & |w| \geqslant \lambda \\ 0, & |w| < \lambda \end{cases} \quad (6)$$

where $\text{sgn}(w)$ denotes the sign function $w$, $w$ is the size of the wavelet coefficients, $w_\lambda$ is the size of the wavelet coefficients after applying the threshold, and $\lambda$ denotes the threshold value.

Hard thresholding is done by making the absolute value of the wavelet coefficients 0 when it is less than the given threshold value and keeping it unchanged when the absolute value of the wavelet coefficients is greater than the threshold value, which is given by the following equation:

$$w_\lambda = \begin{cases} w, & |w| \geqslant \lambda \\ 0, & |w| < \lambda \end{cases} \quad (7)$$

where $w$ is the size of the wavelet coefficients, $w_\lambda$ is the size of the wavelet coefficients after the threshold is applied, and $\lambda$ is expressed as the threshold value.

The wavelet coefficients are reconstructed into the denoised hydroelectric unit signal $T_2$.

### D. Locally Linear Embedding

Locally linear embedding (LLE) is a nonlinear, unsupervised extraction process that produces low-dimensional global coordinates by assuming that data lies on a smooth nonlinear manifold embedded in a high feature space.[14] Imagine we get a sample set $D = x_1, x_2, ... x_i, ..., x_m$, where $i = 1, 2, ..., m$.

Calculate the $k$ nearest neighbors to $x_i$ by Euclidean distance as a metric. The Euclidean distances are as follows:

$$d_{\min} = \sqrt{\sum_{k=1}^{K} (x_i - x_{ik})^2} \quad (8)$$

where $x_{ik}$ is the kth nearest neighbor of $x_i$ and $k = 1, 2, ..., K$.

Then solve for the local covariance matrix, the formula is as follow:

$$Z_i = (x_i - x_j)^T (x_i - x_j) \quad (9)$$

where $Z_i$ represent the local covariance matrix.

After that, find the corresponding vector of weight coefficients, the formula is as follow:

$$W_i = \frac{Z_i^{-1} 1_k}{1_k^T Z_i^{-1} 1_k} \quad (10)$$

where $W_i$ is the *ith* vector of weight coefficients and $1_k$ is a $k$-dimensional all-1 vector.

Afterwards, compute the matrix M, the formula is as follow:

$$M = (I - W)^T (I - W) \quad (11)$$

where the weight coefficient matrix $W$ is formed from the weight coefficient vector $W_i$.

Compute the first $d+1$ eigenvalues of the matrix M and compute the eigenvectors corresponding to these $d+1$ eigenvalues $y_1, y_2, \ldots y_{d+1}$.

Finally, we usually use the second eigenvector to the $(d+1)th$ eigenvector to compose the reduced matrix $D' = (y_2, y_3, \ldots y_{d+1})$.

*E. AdaBoost*

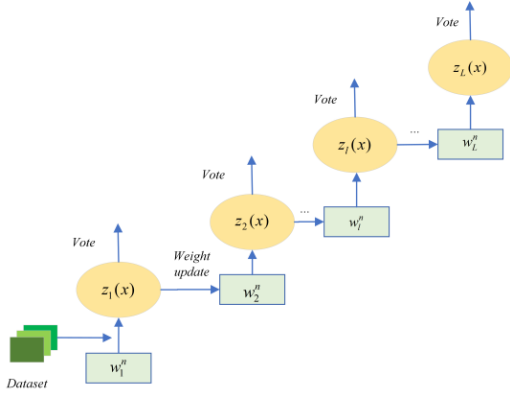

Fig. 1: Schematic diagram of AdaBoost

AdaBoost is adopted in this paper for fault localization. The schematic diagram is shown in Fig. 1. Given a training dataset $T = \{(x_1, y_1), \ldots, (x_h, y_h), \ldots, (x_H, y_H)\}$ where $x_h$ represents the samples, $y_h$ belongs to the category set $\{-1, +1\}$.

Initialize the weight distribution of the training data $H_1 = (w_{11}, w_{12}, \cdots, w_{1k}, \cdots, w_{1H})$.

Assume that there are multiple training samples, each training sample is given the same weights at the beginning:

$$w_{1k} = \frac{1}{H} \quad (12)$$

Multiple rounds of iterations are performed, denoted by $o = 1, 2, \ldots, O$; A weak classifier $G_o(x)$ is obtained by learning using a training dataset with weight distribution $H_o$. Calculate the classification error rate $e_o$ of the weak classifier $G_o(x)$ on the training dataset, the formula is as follow:

$$e_o = \sum_{k=1}^{H} w_{o,k} I(G_o(x_k) \neq y_k) \quad (13)$$

where $e_o$ denotes the classification error rate of o iterations; function $I$ denotes when $I(true) = 1$, $I(false) = 0$. $G_o(x_k)$ denotes the weak classifier corresponding to the *kth* training sample data, and $y_k$ refers to the category corresponding to the *kth* training sample data.

Calculate the weight of $G_o(x)$ in the strong classifier:

$$\alpha_o = \frac{1}{2} \ln \frac{1 - e_o}{e_o} \quad (14)$$

where $\alpha_o$ denotes the importance of $G_o(x)$ in synthesizing a strong classifier, and $e_o$ denotes the classification error rate at $o$ iterations.

Update the weight distribution of the training dataset as:

$$w_{o+1,i} = \frac{w_{o,k}}{Z_o} \exp(-\alpha_o y_k G_o(x_k)) \quad (15)$$

where: $w_{o+1,i}$ denotes the weight of the *ith* training sample data iteration to time $o+1$. The normalization factor is the normalization factor, which makes a probability distribution. $Z_o$ is the normalization factor which enables $H_{o+1}$ to be a probability distribution. The representation of $Z_o$ is as follows:

$$Z_o = \sum_{k=1}^{H} w_{o,k} \exp(-\alpha_o y_k G_o(x_k)) \quad (16)$$

The individual weak classifiers are composed to obtain the final classifier $F(x)$:

$$F(x) = sgn\left(\sum_{o=1}^{O} \alpha_o G_o(x)\right) = sign(f(x)) \quad (17)$$

where $F(x)$ is the final strong classifier, $f(x)$ is a linear combination of weak classifiers, and $sgn$ is the sign-taking function.

III. PROPOSED METHOD

There are three main modules in the proposed SG-WMBDL which are data generation module, data preprocessing module and fault localization module. The flowchart of SG-WMBDL is shown in Fig. 2.

In the data generation module, a small amount of fault sample data from two physical fields are data fused which provides more comprehensive and accurate information thus improving the data quality, and then the fused data are put into SAE and GAN models for fault data generation. This module well solves the problem of very small amount of fault data. The expansion of new fault data can be extended in two major steps. The first step is to fix generator G and maximize discriminator D:

$$\max_{D} L(D) = \max_{D} E_{x \sim P_d} \log[D(x)] + E_{x \sim P_n} \log[1 - D(G(x))] \quad (18)$$

The second step is to fix D and minimize G:

$$\min_{G} L(G) = \min_{G} E_{x \sim P_n} \log[1 - D(G(x))] \quad (19)$$

In the second step, G deceives the discriminator D by minimizing the generation error so that the output of D(G(x)) converges to 1 allowing the output to be minimized.

In the data preprocessing module, noise reduction is on the data based on improved wavelet noise reduction combining soft

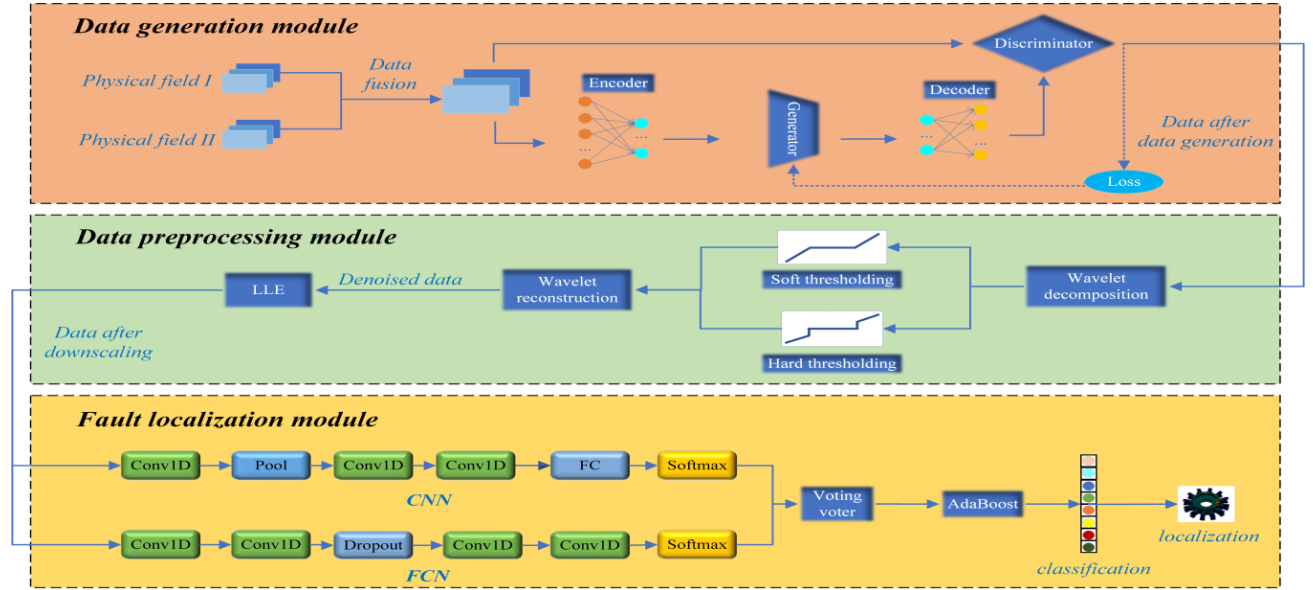
Fig. 2: The flowchart of the proposed SG-WMBDL method

and hard thresholding. Then, the LLE manifold learning method is used to reduce the dimensionality of the hydroelectric unit data, and the ablation experiments prove that the LLE can effectively reduce the dimensionality of the data of the nonlinear hydroelectric unit. This module captures the local features of non-linear, non-smooth hydroelectric unit signals and provides effective noise reduction of hydroelectric unit data. Our innovation is to merge the coefficients from hard and soft thresholding to form the merged wavelet coefficients. The coefficients after hard thresholding are expressed as $C_h = [c_{1,h}, c_{2,h}, \ldots, c_{n,h}]$, the coefficients after soft thresholding are expressed as $C_s = [c_{1,s}, c_{2,s}, \ldots, c_{n,s}]$, where $n$ is the number of coefficients. Then the merged wavelet coefficients are:

$$C = W_h C_h + W_s C_s \quad (20)$$

where $W_h$ is the weight on the hard threshold coefficients, $W_s$ is the weight on the soft threshold coefficients, $C$ is the merged wavelet coefficients.

In the fault localization module, we use the soft voting mechanism effectively to combine the two classifiers of convolutional neural network (CNN) and fully convolutional network (FCN) and realizes the accurate localization of the faults of the hydroelectric unit by introducing the AdaBoost algorithm that dynamically adjusts the weights based on the distribution of the categories. In the AdaBoost algorithm based on the dynamic adjustment of weights of category distribution, we calculate the number of times each category appears in the training set. Then calculate the category weights:

$$W(c) = \frac{L(y_t)}{N \times D(c)} \quad (21)$$

where $c$ is the category label, $W(c)$ is the category weight, $L(y_t)$ is the total number of samples in the training set, $N$ is the total number of categories and $D(c)$ is the initial category distribution. After that the weight distribution of the training dataset is updated again.

To better showcase more design information, here is the pseudocode for SG-WMBDL.

**Algorithm:** SG-WMBDL modeling

1: Collect samples of eight different fault conditions from two physical fields $X_1$, $X_2$.

2: Fuse the data from $X_1$ and $X_2$ into $X_3$ and then sample $X_3$ at intervals then normalize it.

3: Select the parameters, including the sparsity penalty term parameter $\alpha$, the sparsity parameter $\rho$, the threshold value $\lambda$, k-neighbors $k$, learning rate $lr$, number of training epochs, batch size and so on.

4: Put $X_3$ into the data generating module and according to (18)-(19), train $X_3$ to get the expanded data $X_4$.

5: Put $X_4$ into the data preprocessing module and decompose $X_4$ into multiple wavelet coefficients through (5). Then use the merged wavelet coefficients to reconstruct $X_4$ according to (20). After that, perform LLE dimensionality reduction on $X_4$ according to (8)-(11).

6: Apply k-folds cross-validation. Divide $X_4$ into k folds in order. Train k models, each trained using k-1 folds as the training set $X_t$ and the remaining 1 for the validation set $X_v$.

7: Train the fault localization module using training set $X_t$ according to (12)-(17) and (21).

8: Evaluate the validation set $X_v$ results individually on each of the k models.

9: Repeat steps 5-8. Select the appropriate parameters based on the result of the validation set.

10: Train all samples with selected parameters to get a stable SG-WMBDL model.

## IV. CASE STUDY

In this experiment, two operating conditions of the hydroelectric unit are used, corresponding to two physical field distributions, to describe the state of the equipment components from different dimensions. Physical field I represents the total deformation, which refers to the deformation of the structure under load, in millimeters. Physical field II represents the paradigm equivalent force (equivalent force) in megapascals.

The information on the 8 faults is used in this fault localization of the hydroelectric unit, as shown in TABLE I, the total number of fault categories is 8 (8 labels, each label corresponds to faults occurring at different angles), and each category contains 10 samples each of physical field I and physical field II. Where label 0 maps the 0-degree to 45-degree position of the hydroelectric unit, label 1 maps the 45-degree to 90-degree position of the hydroelectric unit, and so on. The precise location of the faults occurring in the hydroelectric unit can facilitate the subsequent fault diagnosis. The original dimensions of the faulty sample are 730301. With the data generation module, each fault data sample is extended from the original 20 to 126 and to simulate the real situation, the training set and test set are set to 2:8 in the following experiments.

extreme gradient boosting (XGBoost) in machine learning, as well as a single neural network CNN and FCN used by the current base classifier, and literature demonstrated the relatively cutting-edge algorithms WDCNN [19] and EVGG [20] as well as CNN-LTSM [21] for comparison of accuracy and precision. Troubleshooting accuracy and precision are tested 10 times averaged for all 8 methods.

From Fig. 3, we can see that the average fault diagnosis accuracy and precision of RF, XGBoost and CNN-LTSM are in the range of 80%-90%, WDCNN and EVGG are in the range of 91%-93%, and CNN, FCN and SG-WMBDL are all able to reach more than 93%, and SG-WMBDL has the highest average fault diagnosis accuracy and precision. The average fault diagnosis accuracy and precision rate of SG-WMBDL are the highest, reaching more than 96%. The experimental results show that for high-dimensional data with nonlinear complexity, the SG-WMBDL method has higher fault localization accuracy and precision than other methods.

TABLE I: Information on the 8 faults

| Category | Sample | Label | Location |
|---|---|---|---|
| Fault 1 | 20 | 0 | 0°-45° |
| Fault 2 | 20 | 1 | 45°-90° |
| Fault 3 | 20 | 2 | 90°-135° |
| Fault 4 | 20 | 3 | 135°-180° |
| Fault 5 | 20 | 4 | 180°-225° |
| Fault 6 | 20 | 5 | 225°-270° |
| Fault 7 | 20 | 6 | 270°-315° |
| Fault 8 | 20 | 7 | 315°-360° |

## A. Performance Comparison with Other Methods

In order to test the performance of the proposed method, the current SG-WMBDL fault localization method for hydroelectric units is compared with random forest (RF) and

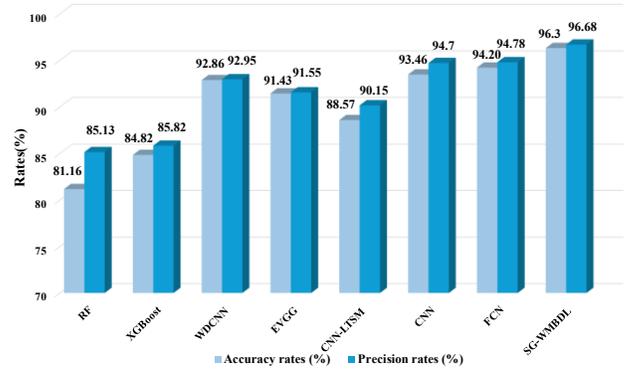

Fig. 3: Accuracy and precision of the 8 methods on the test set

TABLE II: DETAILED ACCURACY RATES (%) AND PRECISION RATES (%) OF DIFFERENT METHODS

| Fault | RF | | XGBoost | | WDCNN | | EVGG | | CNN-LTSM | | CNN | | FCN | | SG-WMBDL | |
|---|---|---|---|---|---|---|---|---|---|---|---|---|---|---|---|---|
| | Acc | Pre | Acc | Pre | Acc | Pre | Acc | Pre | Acc | Pre | Acc | Pre | Acc | Pre | Acc | Pre |
| 1 | 41.18 | 87.50 | 85.71 | 92.31 | **99.03** | 86.44 | **100** | 84.43 | 87.38 | 88.11 | 93.07 | **98.95** | 95.05 | 96.00 | 98.02 | 98.02 |
| 2 | 88.24 | 60.00 | 71.43 | 100 | 77.67 | **100** | 96.12 | 99.05 | 89.32 | 90.12 | 93.07 | **100** | 92.08 | 97.89 | **97.03** | 95.15 |
| 3 | 41.18 | **100** | 92.86 | 92.86 | 95.19 | 92.52 | 81.73 | 94.44 | **100** | 58.43 | 97.03 | 71.53 | **100** | 84.87 | **100** | 96.19 |
| 4 | **100** | 75.00 | 64.29 | 75.00 | 90.38 | **100** | 92.31 | **100** | 90.38 | **100** | 90.20 | 98.92 | 90.20 | **100** | 90.20 | **100** |
| 5 | **100** | 80.95 | 78.57 | 78.57 | 95.19 | 88.39 | 96.15 | **100** | 75.96 | 98.75 | 96.04 | **100** | **100** | 93.52 | 98.02 | **100** |
| 6 | 94.44 | **100** | 92.86 | 81.25 | 90.29 | 96.88 | 92.23 | 87.96 | 86.41 | 89.15 | 90.10 | 97.85 | 90.10 | **100** | 91.09 | **100** |
| 7 | **100** | 77.27 | **100** | 73.68 | 90.38 | **100** | 90.38 | **100** | 90.38 | **100** | 93.14 | 97.94 | 90.20 | **100** | 97.06 | **100** |
| 8 | 82.35 | **100** | 92.86 | 92.86 | 98.06 | 79.53 | 97.09 | 84.75 | 97.09 | 90.09 | 95.05 | 92.31 | 96.04 | 85.84 | **99.01** | 84.03 |

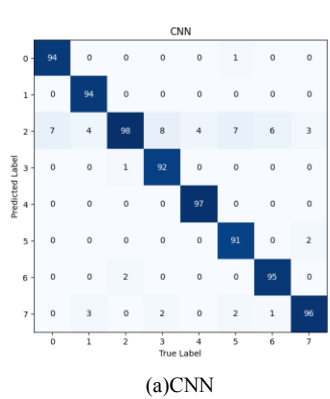
(a) CNN

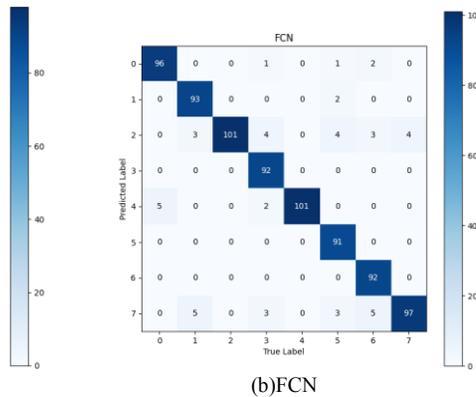
(b) FCN

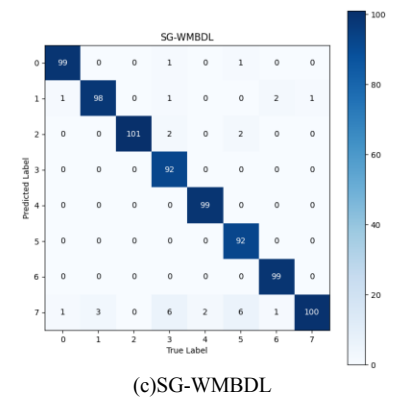
(c) SG-WMBDL

Fig. 4: Confusion matrices of CNN, FCN and SG-WMBDL

More detailed information is shown in TABLE Ⅱ. It is not difficult to find that CNN, FCN and SG-WMBDL using the data generation module can achieve more than 90% fault localization accuracy for these eight faults with good precision. Meanwhile, to make the results clearer and more intuitive, Fig. 4(a), Fig. 4(b) and Fig. 4(c) show the confusion matrices of the base classifiers CNN, FCN and the proposed SG-WMBDL method, respectively. Among them, SG-WMBDL has a much more efficient fault localization. Experimental data suggests that this shows the superiority and practicality of SG-WMBDL.

*B. Ablation Analysis*

TABLE III demonstrates the ablation experiments for each module of the SG-WMBDL with other parameter conditions unchanged. Among them SG-WMBDL (w/o sg) denotes the removal of only the data generation module. SG-WMBDL (w/o w) denotes the SG-WMBDL after removing the wavelet noise reduction based on the combination of improved soft thresholding and hard thresholding and replacing it with the conventional wavelet noise reduction. SG-WMBDL (w/o m) denotes the SG-WMBDL after removing the LLE of manifold learning and replacing it with PCA. SG-WMBDL (w/o b) denotes SG-WMBDL after removing the AdaBoost algorithm that dynamically adjusts the weights according to the category distribution, and replacing it with the traditional AdaBoost.

TABLE III: ABLATION EXPERIMENT OF SG-WMBDL

| Method | Acc /% | Pre /% |
| --- | --- | --- |
| SG-WMBDL (w/o sg) | 84.64 | 85.75 |
| SG-WMBDL (w/o w) | 92.50 | 93.77 |
| SG-WMBDL (w/o m) | 88.93 | 90.92 |
| SG-WMBDL (w/o b) | 94.61 | 94.05 |
| SG-WMBDL | 96.30 | 96.68 |

The results of this ablation experiment can be seen that the method of combining SAE and GAN in the data generation module allows the accuracy and precision to be improved by nearly 12% or so, proving that it can effectively solve the problem of scarce fault data. The improved wavelet noise reduction and LLE dimensionality reduction methods based on the combination of soft and hard thresholds in the second data preprocessing module can improve the accuracy and precision by 4%-8% and the AdaBoost based on the dynamic adjustment of the weights of the category distribution in the fault localization module can improve by about 2%, which proves that it can better extract the non-linear and non-smooth signal features of the hydroelectric unit and accurately locate the faults.

V. CONCLUSION

This paper proposes a new fault localization method for hydroelectric unit, called SG-WMBDL. In the data generation module, a method combining SAE and GAN is adopted. In the data preprocessing module, an improved wavelet noise reduction method based on a combination of soft threshold and hard threshold with LLE is proposed. In the fault localization module, AdaBoost of ensemble learning is used to ensemble CNN and FCN in the neural network to achieve the advantages of both approaches. The SG-WMBDL method is studied and evaluated using hydroelectric unit data. The experimental results show that the SG-WMBDL is able to localize the faults of a hydroelectric unit with higher precision and accuracy than other frontier methods when dealing with a small number of fault samples with non-linear and non-smooth characteristics. These results validate the effectiveness and practicality of our proposed method. In our future work, we plan to further optimize this method and explore the possibility of its application in other types of mechanical fault localization.